\documentclass[a4paper,11pt,dvips]{article}
\usepackage{graphicx}
\usepackage{amsmath}
\usepackage{amssymb}
\usepackage{latexsym}
\usepackage{subcaption}
\usepackage[colorlinks]{hyperref}
\usepackage{color}
\usepackage{enumerate}
\usepackage{enumitem}
\usepackage{cite}
\usepackage{makeidx}
\usepackage{colortbl}

\newcommand{\bra}{\begin{array}}
\newcommand{\era}{\end{array}}
\newcommand{\beq}{\begin{equation}}
\newcommand{\eeq}{\end{equation}}
\newcommand{\bqr}{\begin{eqnarray}}
\newcommand{\eqr}{\end{eqnarray}}

\def\BC{\bb C}
\def\_\BC{\bbi C}

\def\( {\left(}
\def\) {\right)}
\def\no2 {{\textstyle{n\over 2}}}

\newcommand{\om}{\omega}

\newcommand{\app}{\approx}

\newcommand{\da}{\dagger}

\newcommand{\lb}{\label}

\begin{document}

\begin{titlepage}
\setcounter{page}{1}
\renewcommand{\thefootnote}{\fnsymbol{footnote}}

\begin{flushright}
%ucd-tpg 10-01\\
%arXiv:yymm.xxxx
\end{flushright}

\vspace{5mm}
\begin{center}

{\Large \bf {Magnetic Field Effect on Strained Graphene Junctions
}}

\vspace{5mm} {\bf Youness Zahidi}$^{a,b}$,
 {\bf Ilham
Redouani}$^{b}$,
{\bf Ahmed Jellal\footnote{\sf a.jellal@ucd.ac.ma}}$^{b,c}$
and {\bf Hocine Bahlouli}$^{c,d}$

\vspace{5mm}

{$^a$\em EMAFI, Polydisciplinary Faculty, Sultan Moulay Selimane
University, Khouribga, Morocco}

{$^{b}$\em Laboratory of Theoretical Physics,  %Department of Physics,
Faculty of Sciences, Choua\"ib Doukkali University},\\
{\em PO Box 20, 24000 El Jadida, Morocco}

{$^c$\em Saudi Center for Theoretical Physics, Dhahran, Saudi Arabia}

{$^d$\em Physics Department,  King Fahd University
of Petroleum $\&$ Minerals,\\
Dhahran 31261, Saudi Arabia}

%{$^{d}$\em Max Planck Institute for the Physics of Complex Systems,\\
%N\"othnitzer Str. 38, 01187 Dresden} %\\[1em]

\vspace{30mm}

\begin{abstract}

We  %theoretically 
investigate the spin-dependent transport
properties of a ferromagnetic/strained/normal graphene junctions
with central region subjected to
%subject to 
a magnetic field $B$. % in the strain region. 
An
analytical approach, based on  Dirac equation, is implemented
to obtain the eigenstates and eigenvalues of the charge carrier in
three regions. Using the transfer matrix method, we determine the
spin-dependent transmission in the presence of an applied strain
along the armchair  and zigzag  directions of the graphene
sample. We find that the strain remarkably modifies the Landau
levels (LLs) originating from the applied $B$. %magnetic field.
It is shown that the spin up/down energy bands, in the first % left
region, are shifted by the exchange $H_{ex}$ and left the whole spectrum linear
as in the case of pristine graphene. In the central region, the
position of the Dirac point changes due to %because of 
the uniaxial strain
and $B$. %the applied magnetic field in the central region. 
It is also
found that the uniaxial strain in graphene induces a contraction
of the LLs spectra. 
Moreover, the strain  and $B$ modify the shape and position of some peaks  in the transmission
probabilities.

\vspace{3cm}

\noindent PACS numbers:   73.63.-b, 73.23.-b, 72.80.Rj %11.80.-m

\noindent Keywords: Graphene, strain, spin, magnetic
field, Landau levels, transmission.

\end{abstract}
\end{center}
\end{titlepage}

%%%%%%%%%%%%%%%%%%%%%%%%%%%%%%%%%%%%%%%%%%%%%%%%%%%%%%%%%%%%
\section{Introduction}
%%%%%%%%%%%%%%%%%%%%%%%%%%%%%%%%%%%%%%%%%%%%%%%%%%

%Since the monolayer graphene device was first fabricated
%\cite{Novoselov04}, 
Graphene remains among the most fascinating
and attractive subject in modern physics %condensed matter physics 
\cite{Novoselov04}. This is due to
its peculiar physical properties, such as Klein tunneling \cite{
Katsnelson05}, which describes the tunneling behaviors of
relativistic Dirac electrons through a potential barrier. Such an
effect has already been observed experimentally \cite{Stander09}
in graphene systems. Moreover, the dynamics of the charge carriers
obeys a massless Dirac-like equation \cite{Semenoff84}. The 
quasi-particles in graphene exhibit a linear dispersion relation in
the vicinity of the Dirac point as well as % In addition to the special linear
%spectrum, graphene  also exhibits 
many excellent transport
characteristics \cite{Novoselov04,Bolotin08}. Furthermore, the
development of graphene also opens a new and promising route to
nanoelectronics and spintronics because it has a high carrier
mobility and small spin-orbit coupling in addition to a long spin
coherence length that has reached more than one micrometer at room
temperature \cite{3}.

%However i
It is worth mentioning that the major challenge in
designing spintronic devices is the difficulty in generating,
controlling and detecting spin polarized currents. The charge
carriers in graphene are, in general, not spin polarized. However,
the spin polarization is an important concept for novel
spintronics application. Recently, it has been suggested
\cite{Haugen08} that spin polarized carriers can be realized by
depositing a ferromagnetic insulator such as EuO on graphene. This
can induce an exchange proximity interaction \cite{Haugen08,
Semenov07}. Under the influence of exchange field in ferromagnetic
graphene on charge carriers, the current of the system
split into spin up and spin down current components. This effect is due to
the so-called Zeeman effect, which gives rise to spin polarization.
%We note that 
The deposition of ferromagnetic insulator EuO on the
graphene sheet was experimentally realized \cite{Swartz12}. It has
been theoretically predicted that the spin current can be
controlled by gate voltages \cite{Haugen08,Yokoyama08},
magnetic barriers \cite{Dell09, Khodas09} and local strain
\cite{Niu12,Wu14}.

Graphene, a one-atom-thick film, exhibits a truly two-dimensional
nature, which is considered as a flexible membrane. Thus,
it is possible to connect mechanical properties with the
electronic ones. This opens the way to investigate the
interplay between elastic and electronic properties \cite{
Levy10, Pereira09, Soodchomshom11}. The electronic properties of
graphene based nanostructures can be tuned by performing a
deformation on the graphene sample \cite{Haugen08,Ni08,
Mohiuddin09, Huang09}. The applied strain in graphene can be
controlled using different methods \cite{Pereira09}. The
application of a strain on graphene sheet acts on the Dirac
fermions as a pseudomagnetic field \cite{Castro09}.
Experimentally, it has been found that the application  of strain
on graphene nanobubble leads to a huge pseudo-magnetic field
($<300\ T$),  that has never been created in the laboratory
\cite{Levy10}.  Thus, a uniaxial strain larger than 23\% in the
zigzag direction can generate a transport gap in the transmission
\cite{Fogler08, Pereira092}. Importantly, mechanical strains in
graphene can also shift the Dirac points, which causes the Dirac
fermions to have asymmetrical velocity $v_x \neq v_y$
\cite{Soodchomshom11, Choi10}.

We plan to investigate the effect of an applied
magnetic field on the spin transport properties of a ferromagnetic/strained/normal graphene
junctions. The present  system is made up of three regions
where the central one is subjected to the uniaxial strain and $B$.
After writing down the Hamiltonian for each region, 
%Then, 
we solve the corresponding Dirac
equation  to obtain  eigenspinors and eigenvalues.
%of the nergy spectrum
%in each region, separtely. 
Using the boundary conditions together
with transfer matrix approach, we determine the transmission
probabilities in terms of the physical parameters. The strain effects (along armchair and zigzag
directions) on the transmission for zero and non-zero
magnetic field will be analyzed. We show that strain reduces the
transmission along the zigzag direction and increases the transmission
along the armchair direction.

The manuscript is organized as follows. In section 2, we set the
theoretical model 
involving the Hamiltonians describing each region of our system.
These will be used to separately solve Dirac equation to
obtain the solutions of the energy spectrum and in particular the Landau levels in
the central region.
%and solve  the Hamiltonian that describes the
%strained graphene spin along with the stationary solutions of our
%wave equation in different regions of the device. 
In section 3, we
explicitly determine the corresponding transmission probabilities in terms of the physical
parameters. In section 4,
We numerically analyze our results
under suitable configurations of the physical parameters characterizing our system.
%We present and discuss our numerical results under suitable conditions 
 We conclude our work in the final section.
%sum up our findings and main
%conclusion in the final section.

%%%%%%%%%%%%%%%%%%%%%%%%%%%%%%%%%%%%%%%%%%%%%%%%%%%%%%%
\section{Model for a uniaxial strained graphene}
%%%%%%%%%%%%%%%%%%%%%%%%%%%%%%%%%%%%%%%%%%%%%%%%%%%%%%%

\begin{figure}[!ht]
  \centering
\includegraphics[width=5cm,  height=6.7cm]{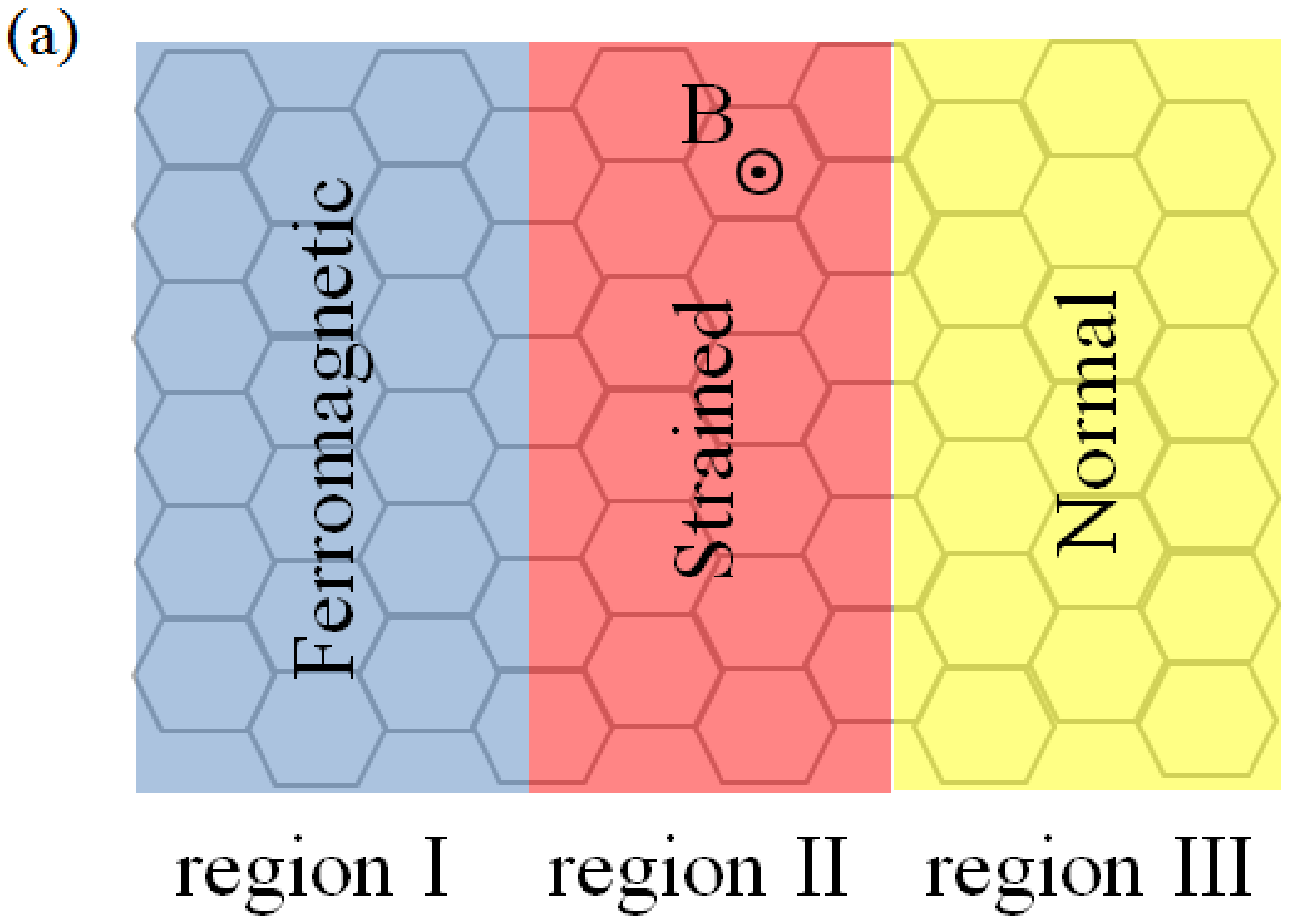} \ \
\includegraphics[width=10.2cm,  height=7cm]{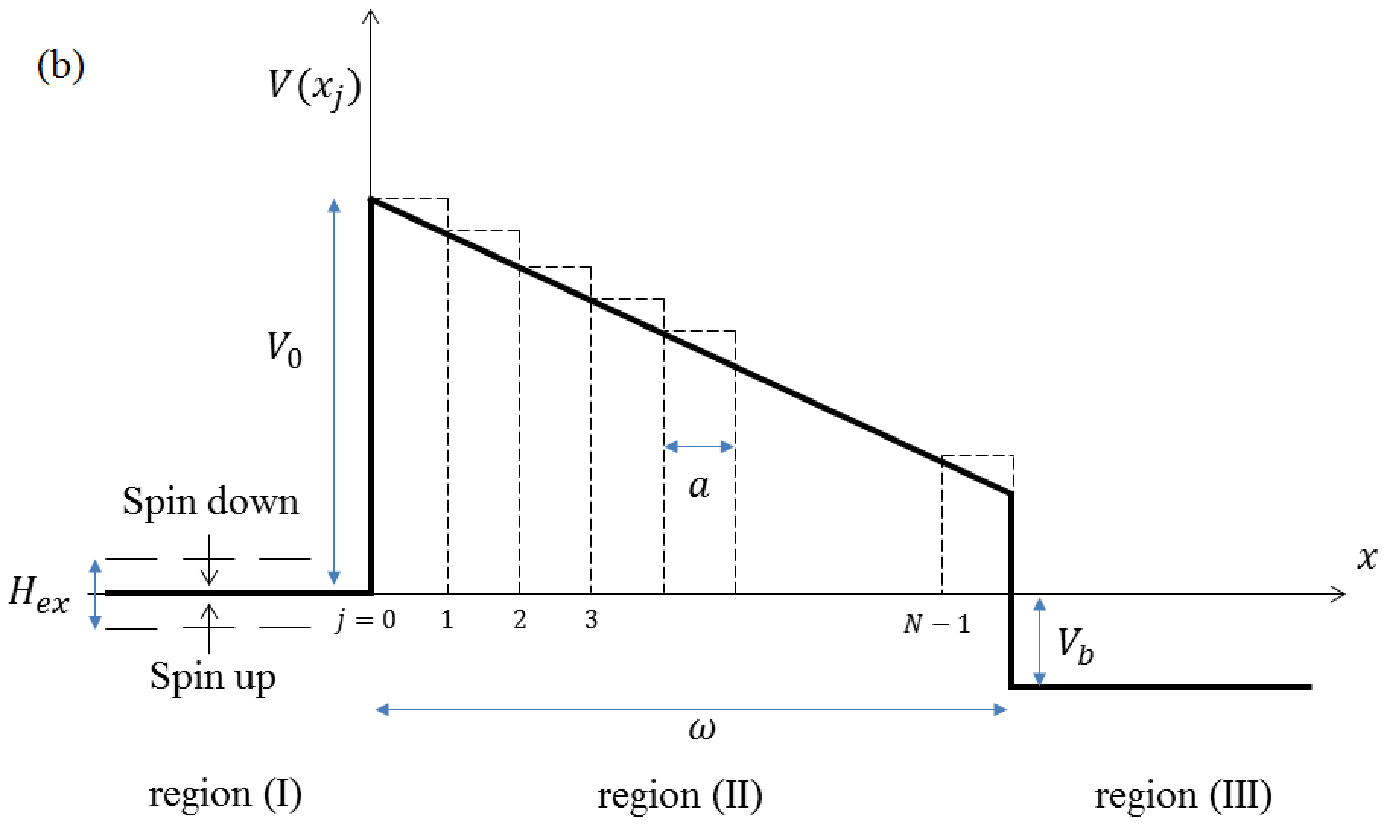}
\caption{\sf{(color online) (a): Schematic illustration of the
ferromagnetic/strained/normal graphene junction with a magnetic
field in the strained region. (b): Potential diagram in different
regions with the exchange field $H_{ex}$, potential $V_0$,
  bias $V_b$,  width of the strained region  $\omega$ and   step of discrimination $a$.}}
\label{1}
\end{figure}

We consider a graphene based system, which is made of three
regions as shown in Figure \ref{1}(a). A
ferromagnetic/strained/normal graphene junction, where a uniaxial
strained graphene sheet of width $\omega$ is sandwiched between a
ferromagnetic and normal metal which are deposited, respectively,
on the left and right regions. %However, by 
Deposing a
ferromagnetic metal, such as EuO, on graphene  can induce an exchange
proximity interaction, which can be treated as an effective
exchange field \cite{Haugen08,Semenov07}. In the central region,
we apply a magnetic field  perpendicular to the graphene layer along
the $z$-direction.

The effective Hamiltonian  describing the quasi-particles  in this
graphene-based system in the presence of a  magnetic field can be written as %given by % \cite{14}
%\beq H=
%\hat{H}_{F,\eta} + \hat{H}_S + \hat{H}_N, \eeq where the three
%parts are given by
\begin{eqnarray}
&&
\hat{H}_{F,\eta} = v_F (\sigma_x \pi_{x_1} + \sigma_y \pi_{y_1})-\eta H_{ex} \sigma_0\\
&& \hat{H}_S =v_F
U^{\dagger}(\theta)[\sigma_x(1-\lambda_x\varepsilon)\pi_{x_2}+
\sigma_y(1-\lambda_y\varepsilon)\pi_{y_2}]{U}(\theta)+(V_0-\beta x)\sigma_0 \\
&& \hat{H}_N =v_F (\sigma_x \pi_{x_3} + \sigma_y \pi_{y_3}) - V_b
\sigma_0
\end{eqnarray}
where $ v_F = {10}^{6}m/s$ is the Fermi velocity, $\sigma_x$ and
$\sigma_y$ are Pauli matrices in pseudospin space, $\sigma_0$ is
the $2\times2$ unit matrix, $\eta = +/-1$ refer to up/down spin
orientation, $H_{ex}$ is the exchange field, ${U}^{+}(\theta) =
$diag$(1,{e}^{ + i \theta})$ is the unitary matrix, $\theta$ is
the angle between the strain direction and the graphene zigzag
direction, $\varepsilon $  is the magnitude of the strain,
$\lambda_x \approx 2.2$ and $\lambda_y \approx$ -0.308. Note
that in the presence of a bias $V_b$ and because of the uniform
in-plane electric field between the left and right electrodes, a
linear voltage drop with a dropping factor of
$\beta=\frac{V_b}{\omega}$, as shown in Figure \ref{1}(b)
\cite{12,13,14}. In addition, the potential barrier $V_0$ will
lift the linear voltage drop completely because of %the potential
$V_0$ is induced by the strain or by an additional gate in the middle
region of width $\omega$. In this work,  we consider 
%It is important to note that the 
a magnitude of the strain 
limited within $20\%$, which means that the energy gap cannot be
opened up \cite{Pereira09}.
The conjugate momenta in the Landau gauge are given by
\begin{eqnarray}
&& \pi_{x_i} = p_x = \hbar q_x = - i\hbar \partial_x\\
&&
\pi_{y_i} = p_y  +  \frac{e}{c} A_{x_i} = \hbar q_y + \frac{e}{c} A_{x_i} =
- i\hbar \partial_y  +  \frac{e}{c} A_{x_i}
\end{eqnarray}
and due to the continuity,
%$l_B \equiv \sqrt{\dfrac{c \hbar}{e B}}$ is the magnetic length.
 %Therefore
 the vector potential in the three regions takes the form
\begin{equation}\label{a}
A_{x_i} =
\dfrac{c\hbar}{e {l_B}^{2}}\left\{
\begin{array}{l}
 0, \qquad x\leq 0\\
 x, \qquad    0<x< \omega\\
\omega,  \qquad x\geq \omega
\end{array}
\right.
\end{equation}
with $l_B \equiv \sqrt{\dfrac{c \hbar}{e B}}$ is the magnetic length and
 $i={\sf I}, {\sf II}, {\sf III}$ labels the three regions. Now we can rewrite the Hamiltonian as
\begin{eqnarray}\label{eq1}
&& \hat{H}_{F,\eta} = v_F (\sigma_x p_{x} + \sigma_y p_{y})-\eta H_{ex} \sigma_0 \\
&& \hat{H}_S =v_F
U^{\dagger}(\theta)\left[\sigma_x\left(1-\lambda_x\varepsilon\right)\hbar
q_x+\sigma_y(1-\lambda_y\varepsilon)
\left(\hbar q_y+\frac{\hbar x}{{l_B}^{2}}\right)\right]U(\theta)+(V_0-\beta x)\sigma_0 \label{eq111}\\
&& \hat{H}_N =v_F \left(\sigma_x p_{x} + \sigma_y
\left(p_{y}+\frac{\hbar w}{{l_B}^{2}}\right)\right) - V_b \sigma_0
\label{eq1111}
\end{eqnarray}
and the components  $q_x$ and $q_y$, measure the wave vector displacements
from the shifted Dirac points \cite{12},  such that
\beq q_D a_0 =\pm \left(k_0
\varepsilon(1+\mu) \cos(3\theta), -k_0 \varepsilon(1+\mu)
\sin(3\theta)\right)
\eeq 
with two quantities  $k_0 \approx 1.6$ and $a_0=0.142\ nm $.

By considering the conservation of the transverse wave vector
$k_y$ and  using (\ref{eq1}-\ref{eq1111}), we can write the eigenspinors
of these quasi-particles,
 moving along the $ \pm x $-directions, in all regions
as $ {\Phi}^{\pm}(x,y) = {\psi}^{\pm}(x) e^{i k_y y}$. Thus, the
solution in the ferromagnetic region {\sf I} can be obtained as
\begin{equation}
  \psi_{\sf I}(x)= \binom{1}{g_\eta {e}^{i \gamma_\eta}}{e}^{i k_\eta x}
  +r_\eta \binom{1}{-g_\eta {e}^{- i \gamma_\eta}}{e}^{- i k_\eta
  x}
\end{equation}
      where  $g_\eta = $sign$ \left(E + \eta H_{ex}\right)$, the angle
      and wave vector
      are given by
      \beq
      \gamma_\eta ={\sin}^{-1} \left(\frac{\hbar v_F ky}{|E + \eta H_{ex}|}\right), \qquad
      k_\eta=\frac{|E + \eta H_{ex}|}{\hbar v_F}\cos\gamma_\eta
      \eeq
     giving rise to the eigenvalues
      \begin{equation}\label{E1_eq2}
      E=g_\eta \hbar v_F \sqrt{ {k_\eta}^2+{k_y}^{2}}- \eta H_{ex}.
      \end{equation}
       In the third region {\sf III} we get the solution
       \begin{equation}
       \psi_{\sf III}(x)= t_\eta \binom{1}{\chi {e}^{i \phi}}{e}^{i k_x x}
\end{equation}
with $\chi=$sign$\left(E + V_{b}\right)$, the angle and wave 
vector read as
\beq
      \phi={\sin}^{-1}
      \left(\frac{\hbar v_F(k_y+\frac{\om}{{l_B}^{2}})}{|E +
      V_{b}|}\right), \qquad k_x =\frac{|E + V_{b}|}{\hbar v_F}\cos\phi
      \eeq
      and the
    corresponding eigenvalues are
    \begin{equation}\lb{E3_eq}
    E=\chi \hbar v_F \sqrt{ {k_x}^2+{\left(k_y+\frac{\om}{{l_B}^{2}}\right)}^{2}}- V_{b}.
    \end{equation}

In the strained region {\sf II} the components of the eigenspinors are no longer plane
waves because of the potential drop.
 To arrive at approximate results, we have split  region 2 into  series of
 reasonably uniform widths $a = \omega/N\gg a_0$ (see Figure \ref{1}(b)), whose potential
 is considered almost constant. In such case, these components %and the wave functions 
 can be considered approximately as
 plane waves with $\alpha_j = j \times a $ ($ j = 0,1,2, \cdots, N $) and $N$ is the number of narrow layers.
The Hamiltonian of  $j^{th}$ narrow layer is given by 
\beq
H_{S,j}=\hbar
v_F{U}^{\dagger}(\theta)\left[\sigma_x(1-\lambda_x\varepsilon)
q_x+ \sigma_y(1-\lambda_y\varepsilon) \left(q_y + \dfrac{
x}{{l_B}^{2}}\right)\right]{U}(\theta) +(V_0-\beta
\alpha_j)\sigma_0
\eeq
 satisfying the eigenvalue equation
 \begin {equation}\label{eq7}
 H_{S,j} {\Phi}_{S,j}(x,y) = E {\Phi}_{S,j}(x,y)
 \end {equation}
 which can explicitly be written  as
\begin{eqnarray}\label{eq88}
&&
  -i\hbar v_F(1-\lambda_x\varepsilon) \left(\partial_x +
  \frac{1-\lambda_y\varepsilon}{1-\lambda_x\varepsilon}
  \left(k_y- q_{D_y} + \frac{x}{{l_B}^{2}} \right)\right){{e}^{-i \theta} \psi}^{B}_{\sf II}(x) =
  \left( E + \beta \alpha_j - V_0 \right){\psi}^{A}_{\sf II}(x)\\
  &&
  i\hbar v_F (1-\lambda_x\varepsilon) \left( -\partial_x +
  \frac{1-\lambda_y\varepsilon}{1-\lambda_x\varepsilon}
  \left(k_y- q_{D_y} + \frac{x}{{l_B}^{2}} \right) \right)
  { {e}^{+i \theta}\psi}^{A}_{\sf II}(x)=( E + \beta \alpha_j - V_0 ){\psi}^{B}_{\sf II}(x). \label{eq888}
\end{eqnarray}
Making use of %With
the variable change
$X=l_B \sqrt{\alpha}\left(k_y-q_{D_y} + \dfrac{x}{{l_B}^{2}}\right)$ and 
strain parameter   $\alpha = \frac{1-\lambda_y\varepsilon}{1-\lambda_x\varepsilon}$
to map (\ref{eq88}-\ref{eq888}) as 
\begin{eqnarray}\label{eq8}
&&
  -i\hbar v_F(1-\lambda_x\varepsilon)\frac{\sqrt{\alpha}}{l_B}(\partial_X + X)
  {{e}^{-i \theta} \psi}^{B}_{\sf II}(x) =( E + \beta \alpha_j - V_0 ){\psi}^{A}_{\sf II}(x)\\
  &&
  i\hbar v_F(1-\lambda_x\varepsilon)\frac{\sqrt{\alpha}}{l_B}
  ( -\partial_X + X){ {e}^{+i \theta}\psi}^{A}_{\sf II}(x)=( E + \beta \alpha_j - V_0
  ){\psi}^{B}_{\sf II}(x)
\end{eqnarray}
which can be  solved by
introducing the usual bosonic operators
\beq
{a}^{\dagger} = \frac{ 1}{\sqrt{2}} \left(-\partial_X + X \right), \qquad
 a = \frac{ 1}{\sqrt{2}} \left( \partial_X + X
 \right)
 \eeq
 satisfying the commutation relation
 $[a,{a}^{\dagger}]=\mathbb{I}$. Thus, we write
\begin{eqnarray}\label{eq9}
 &&
 -i\hbar v_F \frac{\sqrt{2\alpha}}{l_B}(1-\lambda_x\varepsilon) a
 { {e}^{-i \theta}\psi}^{B}_{\sf II}(x)
 =( E + \beta \alpha_j - V_0 ){\psi}^{A}_{\sf II}(x) \\
 &&
  i\hbar v_F \frac{\sqrt{2\alpha}}{l_B}(1-\lambda_x\varepsilon) {a}^{\da}
  {{e}^{+i \theta} \psi}^{A}_{\sf II}(x)=( E + \beta \alpha_j - V_0 ){\psi}^{B}_{\sf II}(x)
  \lb{15.2}
\end{eqnarray}
giving rise to the second order differential equation for ${\psi}^{A}_{\sf II}(x)$
\begin{equation}\lb{hosc}
\frac{2 \alpha}{{l_B}^{2}}{(1-\lambda_x\varepsilon)}^{2} a
{a}^{\dagger} {\psi}^{A}_{2}(x) = {\left(\frac{ E + \beta \alpha_j
- V_0}{\hbar v_F} \right)}^{2}{ \psi}^{A}_{\sf II}(x).
\end{equation}
It  is clear that \eqref{hosc} is similar to that describing
the harmonic oscillator in one dimension
and therefore we can identify
${\psi}^{A}_{\sf II}$
to be a harmonic oscillator
 eigenstate
\begin{equation}\label{16}
{\psi}^{A}_{\sf II}\sim \mid n-1\rangle
\end{equation}
associated to the eigenvalues
  \begin{equation}\lb{eq.E2-2}
 E_{n,j} = {\pm} \hbar v_F \frac{1-\lambda_x\varepsilon}{l_B} \sqrt{2\alpha n } + V_0 - \beta\alpha_j
 \end{equation}
 and the second spinor component can be obtained by
injecting \eqref{16} into \eqref{15.2} %to obtain
\begin{equation}
{\psi}^{B}_{\sf II} = i \frac{\sqrt{2\alpha}{e}^{+i
\theta}}{l_B\left(\frac{ E + \beta\alpha_j - V_0} {\hbar
v_F}\right )}(1-\lambda_x\varepsilon) {a}^{\da}\mid n-1\rangle.
\end{equation}
 It is convenient to work with  the parabolic cylindrical functions
\beq  
D_{n}(x) = {2}^{- \frac{n}{2}} {e}^{- \frac{{z}^{2}}{4}}
H_{n} \left(\frac{x}{\sqrt{2}}\right) 
\eeq 
and express the solution
in the central region as 
\beq 
\psi_{\sf II}(x_j)=a_j
\psi^{+}_{\sf II}(x_j)+b_j \psi^{-}_{\sf II}(x_j)
\eeq 
where the components are given by
\begin{equation}
\label{eq3} {{\psi^{\pm}}_{\sf II}}(x_j)=
\begin{pmatrix}
 D_{{n-1}} \left[\pm \sqrt{2 \alpha}\left(l_B (k_y- q_{D_y}) +
 \frac{x_j}{l_B}\right) \right]\\\\
 \pm i \Omega_{j} {e}^{+i \theta} D_{n} \left[\pm \sqrt{2 \alpha}\left(l_B (k_y- q_{D_y})
 + \frac{x_j}{l_B}\right)\right]\\
\end{pmatrix}
\end{equation}
with $(a_j, b_j)$ are two constants and  we have set  the quantity
\begin{equation}\label{22}
 \Omega_{j}=\frac{\hbar v_F \sqrt{2 \alpha }(1-\lambda_x\varepsilon)}{l_B \mid E_{n,j}
 + \beta\alpha_j - V_0\mid}.
\end{equation}
It is clearly seen that the above solutions show strong dependency on
the strain and magnetic field effects. %These will offer different posibilitiess
In the forthcoming analysis, we will study their influence
%see the influence of such effects 
on the transmission of the charge carriers
trough the considered potential barrier.

%%%%%%%%%%%%%%%%%%%%%%%%%%%%%%%%%%%%%%%%%%%%
\section{Transmission probabilities}
%%%%%%%%%%%%%%%%%%%%%%%%%%%%%%%%%%%%%%%%%%

We analyze the strain effect on the transmission probability
in ferromagnetic/strained/normal graphene junction under magnetic
field.
To determine the transmission and reflection probabilities, we use the 
corresponding current densities
\beq 
T_{\eta} = \left|\frac {j_{{\sf tra}}^{\eta}}{j_{{\sf
inc}}^{\eta}}\right|, \qquad R_{\eta} = \left|\frac {j_{{\sf
ref}}^{\eta}}{j_{{\sf inc}}^{\eta}}\right| 
\eeq 
where $(j_{\sf
inc}^{\eta}, j_{\sf ref}^{\eta}, j_{\sf tran}^{\eta})$ are,
respectively, the probability current density of the incident,
reflected and transmitted waves. For a relativistic quasi-particle
propagating along the positive $x$-direction, the current density
is given by
\begin{equation}
 J= e v_F \overline{{\psi}} \sigma_x \psi
\end{equation}
and therefore we derive the results
%which gives the following results for the incident, reflected and transmitted components
\begin{eqnarray}
&& j_{\sf inc}^{\eta} = 2 e v_F \frac{k_\eta}{E+\eta H_{ex}}\\
&&
j_{\sf ref}^{\eta} = - 2 e v_F \frac{k_\eta}{E+\eta H_{ex}} r_\eta^{*} r_\eta\\
&& j_{\sf tra}^{\eta} = 2 e v_F \frac{k_x}{E+V_b}t_\eta^{*}
t_\eta
\end{eqnarray}
%giving rise to 
%allowing to obtain
leading to
the probabilities
\begin{eqnarray}\label{29}
&& T_{\eta} = \left|\frac {j_{\sf tra}^{\eta}}{j_{\sf inc}^{\eta}}\right| = \frac{k_x \left(E+\eta H_{ex}\right)}
{k_\eta\left(E+V_b\right)} |t_\eta|^2\\
&&
R_{\eta} = \left|\frac {j_{\sf ref}^{\eta}}{j_{\sf inc}^{\eta}}\right|={|r_\eta|}^{2}.
\end{eqnarray}
To go further, %It is clear that
we need  to determine the transmission $t_\eta$ and
reflection $r_\eta$ coefficients. Indeed, since the
% Finally for all incident
quasi-particles have normalized probability densities, then
 we can write the eigenspinors corresponding
 to the three regions as % total wave functions in the corresponding regions\cite{25}
 \begin{equation}\label{totp}
 {\Phi}(x,y)=\left\{
\begin{array}{l}
 ~~ {U}^{+}(\theta){\psi}_{\sf I}(x) {e}^{  i k_y y}, ~~~~~~~~~~~x\leq 0~~~~\\
 ~~ {U}^{+}(\theta){\psi}_{\sf II}(x) {e}^{  i k_y y}, ~~~~~~~~~~~ 0<x< \om~~~~\\
 ~~ {U}^{+}(\theta){\psi}_{\sf III}(x){e}^{  i k_y y}, ~~~~~~~~~~  ~~x\geq \om~~~~
\end{array}
\right. \end{equation}
where different eigenspinors in $x$-direction take the forms
\begin{eqnarray}\label{25}
\psi_{\sf I}(x) &=& \left( \begin{array}{c} 1  \\
 g_\eta {e}^{i \gamma_\eta} \\
 \end{array}
  \right){e}^{+ i k_\eta x} +r_\eta \left(\begin{array}{c}
                                                                                 1  \\
                                                                                 -g_\eta {e}^{-i \gamma_\eta} \\
                                                                               \end{array}
                                                                             \right){e}^{- i k_\eta x}\\
\psi_{\sf II}(x_j)&=& a_j \begin{pmatrix}
 D_{{n-1}} \left[+ \sqrt{2 \alpha}\left(l_B (k_y- q_{D_y}) + \frac{x_j}{l_B}\right) \right]\\
 + i \Omega_{j} {e}^{+i \theta} D_{n} \left[+\sqrt{2 \alpha}\left(l_B (k_y- q_{D_y})
 + \frac{x_j}{l_B}\right)\right]\\
\end{pmatrix} \nonumber \\
&&+ b_j \begin{pmatrix}
 D_{{n-1}} \left[-\sqrt{2 \alpha}\left(l_B (k_y- q_{D_y}) + \frac{x_j}{l_B}\right) \right]\\
 - i \Omega_{j} {e}^{+i \theta} D_{n} \left[-\sqrt{2 \alpha}\left(l_B (k_y- q_{D_y}) +
 \frac{x_j}{l_B}\right)\right]\\
\end{pmatrix} \\
\psi_{\sf III}(x) &=& t_\eta\left(\begin{array}{c}
                                                                                 1  \\
                                                                                 \chi {e}^{+ i \phi} \\
                                                                               \end{array}
                                                                             \right){e}^{+ i k_x x}
\end{eqnarray}
with $j= 0,1,2, \cdots, N$ and $N$ is the number of
narrow layers. 
% are the transmission and reflection
%coefficients, respectively \cite{13,14,Pereira09,16},  $a_j$
%and $b_j$ are unknown complex coefficients that both contain a
%factor of the strain-changed group velocity \cite{Haugen08,18,19}.
%Remember that  
To determine the coefficients  $r_\eta$, $t_\eta$, $a_j$ and $b_j$ using the
continuity equations, we define the shorthand notations
 \begin{eqnarray}
&&
{\eta}^{\pm}_j(x_{j}) =  D_{{n-1}} \left[\pm \sqrt{2 \alpha}\left(l_B (k_y- q_{D_y}) +
\frac{x_j}{l_B}\right) \right]\\
&&
{\delta}^{\pm}_j (x_{j})=\pm i \Omega_{j} D_{n} \left[\pm \sqrt{2 \alpha}\left(l_B (k_y- q_{D_y}) +
\frac{x_j}{l_B}\right)\right].
 \end{eqnarray}
The continuity of the eigenspinors \eqref{totp}  at each %junction
interface   results in a set of equations.
Indeed, at the $ x_{j}$ $(j=0, 1 , 2, \cdots, N)$, we have,
respectively, for $x=0$, $x=a$, $x=2a$ $\cdots $ and $x=Na=\om$
\begin{eqnarray}
 && \left\{
\begin{array}{l}
1+r_\eta= {\eta}^{+}_0 a_0 +  {\eta}^{-}_0 b_0\\
g_\eta {e}^{i \gamma_\eta}-g_\eta r_\eta {e}^{- i \gamma_\eta}=i  {\delta}^{+}_0 a_0 - i {\delta}^{-}_0 b_0
\end{array}
\right.\\
&& \left\{
\begin{array}{l}
{\eta}^{+}_0(a) a_0 +  {\eta}^{-}_0(a) b_0= {\eta}^{+}_1 (a) a_1 +  {\eta}^{-}_1 (a) b_1\\
i {\delta}^{+}_0(a) a_0 -i  {\delta}^{-}_0(a) b_0= i {\delta}^{+}_1 a_1 -i {\delta}^{-}_1 (a) b_1
\end{array}
\right.\\
&&
 \left\{
\begin{array}{l}
{\eta}^{+}_1(2a) a_1 +  {\eta}^{-}_1(2a) b_1= {\eta}^{+}_2 (2a) a_2 +  {\eta}^{-}_2 (2a) b_2\\
i {\delta}^{+}_1(2a) a_1 -i{\delta}^{-}_1(2a) b_1= i {\delta}^{+}_2 (2a) a_2 -i  {\delta}^{-}_2 (2a) b_2
\end{array}
\right.\\
&&
\begin{array}{l}\nonumber
\vdots
\end{array} \\
 &&
  \left\{
\begin{array}{l}
 {\eta}^{+}_{N-1}(Na) a_{N-1} +  {\eta}^{-}_{N-1} (Na) b_{N-1}= t_\eta {e}^{i k_x w}\\
i {\delta}^{+}_{N-1} (Na) a_{N-1} - i {\delta}^{-}_{N-1} (Na) b_{N-1}= \chi {e}^{i \varphi}
t_\eta {e}^{i k_x w}
\end{array}
\right.
\end{eqnarray}
%This set of equations
which
can be cast as
\begin{eqnarray}
\left(
                 \begin{array}{l}
                   1 \\
                   r_\eta \\
                 \end{array}
               \right) &=&
{\left(\begin{array}{ll}
    1 &  1 \\
    g_\eta{e}^{i \gamma_\eta} & - g_\eta{e}^{-i \gamma_\eta} \\
  \end{array}
\right)}^{-1}\left(\begin{array}{ll}
    {\eta}^{+}_0(0) &  {\eta}^{-}_0(0) \\
    {i \delta}^{+}_0 (0) & -i{\delta}^{-}_0 (0) \\
  \end{array}
\right){\left(\begin{array}{ll}
    {\eta}^{+}_0(a) &  {\eta}^{-}_0(a) \\
    {i \delta}^{+}_0 (a) & -i{\delta}^{-}_0 (a) \\
  \end{array}
\right)}^{-1}\nonumber\\
&&
\left(\begin{array}{ll}
    {\eta}^{+}_1(a) &  {\eta}^{-}_1(a) \\
    {i\delta}^{+}_1 (a) & {-i \delta}^{-}_1 (a) \\
  \end{array}
\right)
{\left(\begin{array}{ll}
    {\eta}^{+}_1(2a) &  {\eta}^{-}_1(2a) \\
    {i\delta}^{+}_1 (2a) & {-i \delta}^{-}_1 (2a) \\
  \end{array}
\right)}^{-1}
\left(\begin{array}{ll}
    {\eta}^{+}_2(2a) &  {\eta}^{-}_2(2a) \\
    {i \delta}^{+}_2 (2a) & {-i \delta}^{-}_2 (2a) \\
  \end{array}
\right)\nonumber\\
&&
{\left(\begin{array}{ll}
    {\eta}^{+}_2(3a) &  {\eta}^{-}_2(3a) \\
    {i \delta}^{+}_2 (3a) & {-i \delta}^{-}_2 (3a) \\
  \end{array}
\right)}^{-1} \cdots {\left(\begin{array}{ll}
    {\eta}^{+}_{N-1}(Na) &  {\eta}^{-}_{N-1}(Na) \\
    {i \delta}^{+}_{N-1} (Na) & {-i \delta}^{-}_{N-1} (Na) \\
  \end{array}
\right)}^{-1}\nonumber \\
&&
\left(\begin{array}{ll}
     e^{i k_x w} &  e^{-i k_x w} \\
    \chi e^{i k_x w}e^{i\phi} & -\chi e^{-i k_x w} e^{-i\phi} \\
  \end{array}
\right)
%\nonumber\\
 % &&
\left(
                 \begin{array}{l}
                   t_\eta \\
                   0 \\
                 \end{array}
               \right)
               \end{eqnarray}
 or equivalently to
\begin{equation}\lb{47}
 \left(
                 \begin{array}{l}
                   1 \\
                   r_\eta \\
                 \end{array}
               \right)=
{\left(\begin{array}{ll}
    1 &  1 \\
    g_\eta{e}^{i \gamma_\eta} & - g_\eta{e}^{-i \gamma_\eta} \\
  \end{array}
\right)}^{-1}{\prod }^{N-1}_{j=0}\tau_j {\sigma_j}^{-1}\left(\begin{array}{ll}
     e^{i k_x w} &  e^{-i k_x w} \\
    \chi e^{i k_x w}e^{i\phi} & -\chi e^{-i k_x w} e^{-i\phi} \\
  \end{array}
\right)\left(
                 \begin{array}{l}
                   t_\eta \\
                   0 \\
                 \end{array}
               \right)
\end{equation}
where we have set the quantities
\begin{equation}
  \tau_j=\left(\begin{array}{ll}
    {\eta}^{+}_{j}(ja) &  {\eta}^{-}_{j}(ja) \\
    {\delta}^{+}_{j} (ja) & {\delta}^{-}_{j} (ja) \\
  \end{array}
\right), \qquad \sigma_j=\left(\begin{array}{ll}
    {\eta}^{+}_{j}((j+1)a) &  {\eta}^{-}_{j}((j+1)a) \\
    {\delta}^{+}_{j} ((j+1)a) & {\delta}^{-}_{j} ((j+1)a) \\
  \end{array}
\right)
\end{equation}
%with $j=0,1,2, \cdots, N.$
We show that \eqref{47} can be written in compact form
as %gives the relation
\begin{equation}
 \left(
                 \begin{array}{l}
                   1 \\
                   r_\eta \\
                 \end{array}
               \right)=
\left(\begin{array}{ll}
    {M}^{\eta}_{11} & {M}^{\eta}_{12} \\
    {M}^{\eta}_{21} & {M}^{\eta}_{22} \\
  \end{array}
\right)\left(
                 \begin{array}{l}
                   t_\eta  \\
                   0 \\
                 \end{array}
               \right)
\end{equation}
and the coefficients are given by
%To solve this system of equations :\par
\begin{equation}\label{39}
t_\eta = \frac{1}{{M}^{\eta}_{11}}, \qquad 
r_\eta = \frac{{M}^{\eta}_{21}}{{M}^{\eta}_{11}}
\end{equation}
leading to the transmission and reflection probabilities
\begin{eqnarray}\label{40}
 T_{\eta} = \frac{k_x \left(E +\eta H_{ex}\right)}{k_{\eta}
\left(E+V_b\right)}
\frac{1}{{\left({M}^{\eta}_{11}\right)}^{*}{M}^{\eta}_{11}}, \qquad 
R_{\eta} = {\left(\frac{{M}^{\eta}_{21}}{{M}^{\eta}_{11}}\right)}^{*} \frac{{M}^{\eta}_{21}}{{M}^{\eta}_{11}}.
\end{eqnarray}
%where we have replaced   %$\eta=\pm1$ the spin state $\eta$ by $\pm$.
The obtained  results so far will be analyzed  numerically
and discussed using some suitable selections of the physical parameters
characterizing our system.
%\newpage

%%%%%%%%%%%%%%%%%%%%%%%%%%%%%%%%%%%%%
\section{Results and discussions}
%%%%%%%%%%%%%%%%%%%%%%%%%%%%%%%%%%%%

We will investigate the physical behavior of our system using a
numerical implementation of the %result obtained
previous theoretical model to compute the energy {spectrum, transmission
probability and {examine} the  magnetic field and strain effects.
 First, we define nanostructures derived from
graphene, called nanorrubans of graphene. These are unidimensional
graphene-based structures as shown in Figure \ref{1}, which %and effectively these two nanoribbons
possess remarkable properties especially under the effects of the
electric and magnetic fields. For this, we consider two typical strain directions
 including the zigzag $(\theta = 0)$ and armchair $(\theta=
\frac{\pi}{2})$ directions.
%as presented in Figure \ref{1}.

Before proceeding to discuss spin-dependent transmission, we will
study the respective energy spectrum in each region of the
considered device. In the left region ({\sf I}) the dispersion relation is
described by \eqref{E1_eq2} and the energy bands are illustrated
in Figure \ref{reg1Eky}(a). %From this figure, we can
We clearly see
that the spin up and down bands are shifted by an amount equal to
the exchange energy $H_{ex}$ and left the whole spectrum linear as
in the case of pristine graphene. %However, 
The dispersion
relation, in the right region ({\sf III}) is described by \eqref{E3_eq} and in
Figure \ref{reg1Eky}(b) we show the energy spectrum 
%in the right region of the device 
for different values of the magnetic filed $B$.
We observe that for zero magnetic field, the energy bands are
linear and similar to {those} in pristine graphene, but shifted down
by $V_b$ due to the bias. By increasing $B$, %the magnetic field, we can
we see that the energy spectrum is still linear, but
the Dirac point is shifted and such shift %It is important to note that the Dirac point is shifted 
is as a consequence of the
uniaxial strain applied in the central region ({\sf II}) as already found in
\cite{12}.

\begin{figure}[!ht]
\centering
\includegraphics[width=8.3cm]{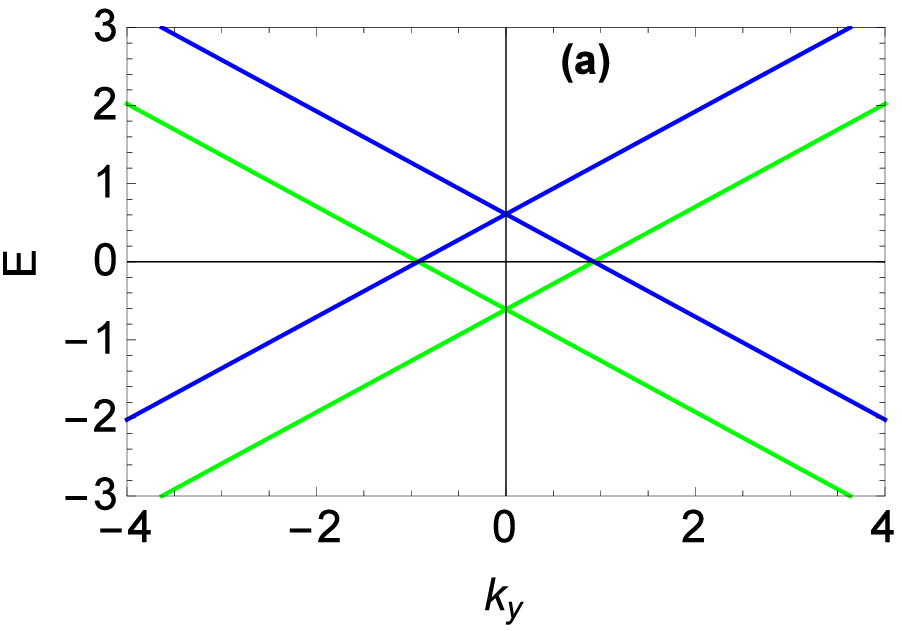}
\includegraphics[width=8.3cm]{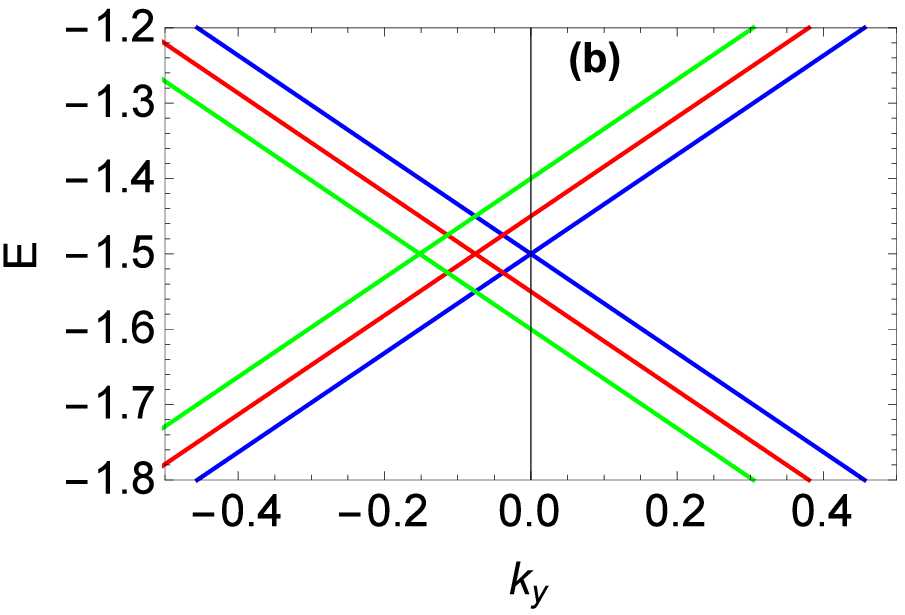}
\caption{ {\sf (color online) {(a): Energy bands of region I as a
function of the momentum $k_y$ with $H_{ex}=0.61\ meV$, green
and blue lines correspond, respectively, to upper $(\eta=+1)$ and
lower $(\eta=-1)$ spin. (b): Energy bands of region
III as a function of the momentum $k_y$ with $V_b=1.5\ meV$,
$\omega=1\ nm$,  blue, red and green lines correspond to $ B=0\
T$,$ B=50\ T$ and $ B=100\ T$, respectively}}}\label{reg1Eky}
\end{figure}

In the central region ({\sf II}), the story is completely different because
besides the strain effect we have also a perpendicular magnetic field that
leaves the energy spectrum quantized, which results in the so-called 
Landau levels (LLs), see \eqref{eq.E2-2}. The evolution of the LLs spectra as 
a function of the strain $\varepsilon$ is shown in Figure \ref{Eeps-B}(a). 
{To thoroughly examine the effect of strain on LLs}, 
 we rewrite %the LLs spectra given in
\eqref{eq.E2-2} 
as 
\beq \lb{Ereno} E_{n,j} = {\pm} \hbar \omega_c
\xi  \sqrt{ n } + V_0 - \beta\alpha_j, \qquad n\in \mathbb{N}, \qquad
j=0,1,2, \cdots, N
\eeq 
where
$\omega_c=v_F\frac{\sqrt{2}}{l_B}$ is the cyclotron frequency and
$\xi=\sqrt{(1-\lambda_x \varepsilon)(1-\lambda_y \varepsilon)}$ is a parameter of strain.
Moreover, one sees that the first term in \eqref{Ereno} is similar
to that corresponding to pristine graphene in magnetic field $B$ apart from a
renormalization of the Fermi velocity $v^{*}_{F}=\xi v_{F}$
\cite{Goerbig08}.  Furthermore, $\xi$ is considered
as a parameter to measure the contraction or the expansion
($\xi<1$ or $\xi>1$) of the LLs spectra under the same magnetic
field. It should be noticed that the limiting case $\xi=1$
corresponds actually to pristine graphene.

In Figure \ref{Eeps-xi}, we show the evolution of the $\xi$
parameter as a function of the strain $\varepsilon$. It is clearly
shown that, $\xi=1$ for $\varepsilon=0$ ($v^{*}_{F}=
v_{F}$), which corresponds to pristine graphene and confirm what we
have already mentioned above. By increasing the strain
$\varepsilon$, the parameter $\xi$ decreases \cite{Goerbig08} and
automatically $v^{*}_{F}$ decreases. From this Figure, it is
clearly seen that whatever the strain deformation (under 20 \%)
we have $\xi<1$, which means that the LLs are contracted {compared} to the pristine graphene.

\begin{figure}[!ht]
\centering
\includegraphics[width=9cm]{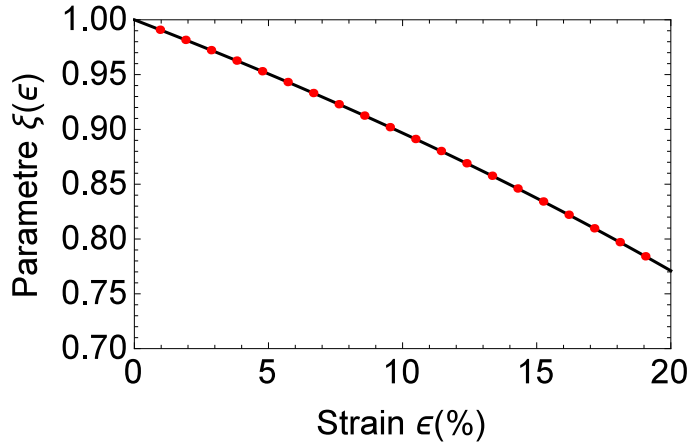}
\caption{ {\sf (color online) {Evolution of the  parameter $\xi$ for strained
graphene as a function of the strain
$\varepsilon$.}}}\label{Eeps-xi}
\end{figure}

From Figure \ref{Eeps-B}(a), we clearly see that under the
same magnetic field, the energy decreases by increasing the
magnitude $\varepsilon$ of the uniaxial strain. Moreover, the distance between
the LLs decreases as well  by increasing $\varepsilon$. Thus, the LLs
spectra are contracted, which means that $\xi<1$, as compared to
pristine graphene ($\xi=1$). This can be explained by the fact
that the induced uniaxial strain affects the cyclotron orbital
motion. It is important to mention that this result is similar to
that obtained in \cite{Betancur15}. To show the effect of the
applied magnetic field, we plot the energy as a function of the
magnetic field in Figure \ref{Eeps-B}(b). Note that the LLs
spectra \eqref{eq.E2-2} depend on the square root of
both the level index $n$ and the magnetic field $B$. From this
Figure, we can show that for zero magnetic field the energy is
not degenerate and  by increasing $B$ the LLs increase and become degenerate.
%as well as
%increase.
%. Note
%that by increasing $B$, the LLs increases.

\begin{figure}[!ht]
\centering
\includegraphics[width=8.3cm]{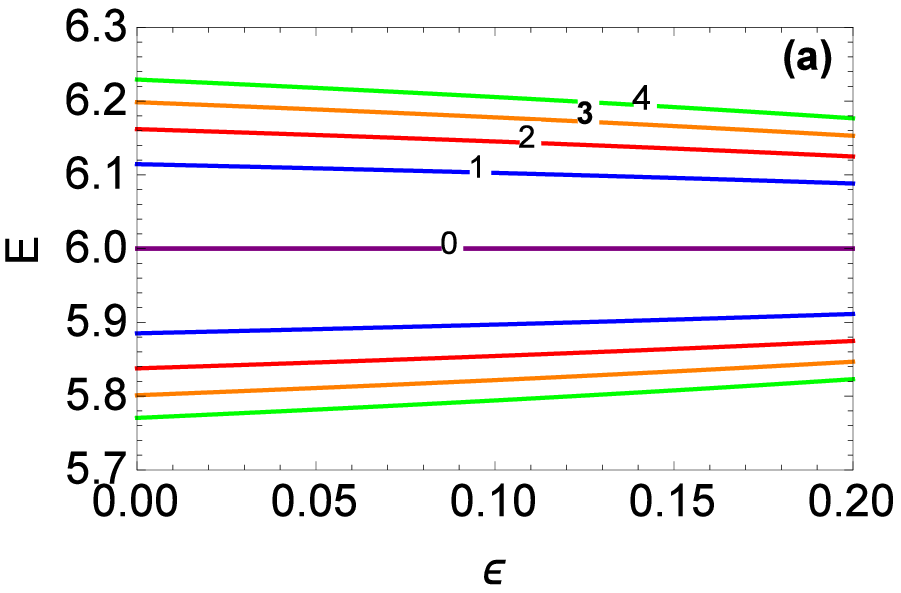}\ \
\includegraphics[width=8.3cm]{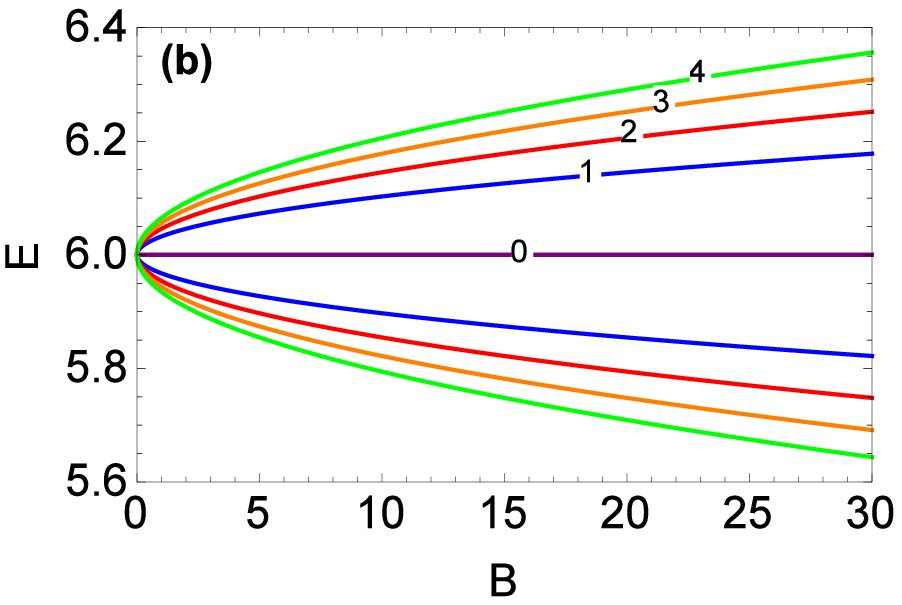}
\caption{ {\sf (color online) {(a): Energy bands of region {\sf II} as
a function of the strain $\varepsilon$ with $B=10\ T$, $\omega=0.1
\ nm$ and $n=0, \cdots, 4$.  (b): Energy bands of
region {\sf II} as a function of the magnetic field $B$ with
$\omega=0.2 \ nm$, $\varepsilon=0.1$, $V_b=6$ and $n=0, \cdots,
4$. }}}\label{Eeps-B}
\end{figure}

\begin{figure}[!ht]
\centering
\includegraphics[width=8.3cm]{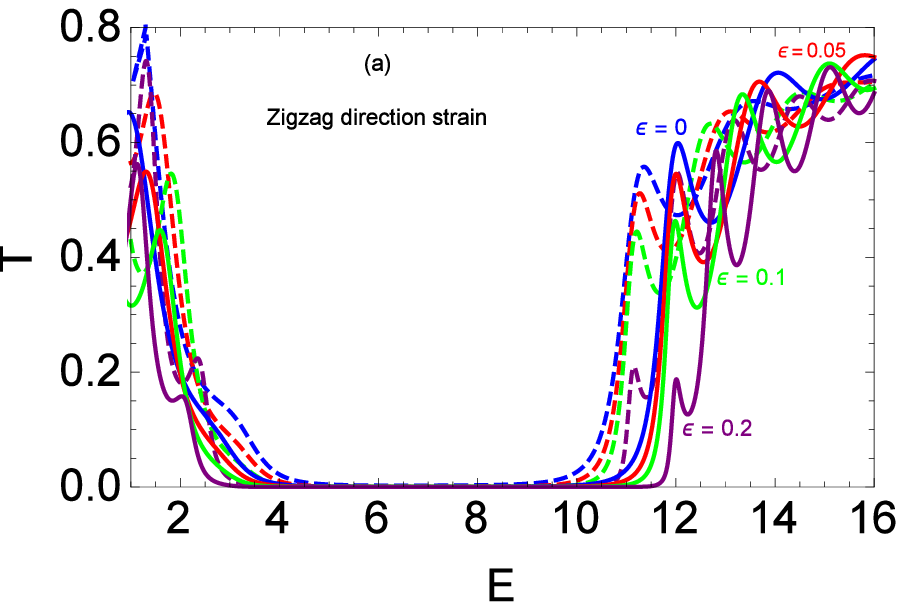}
\includegraphics[width=8.3cm]{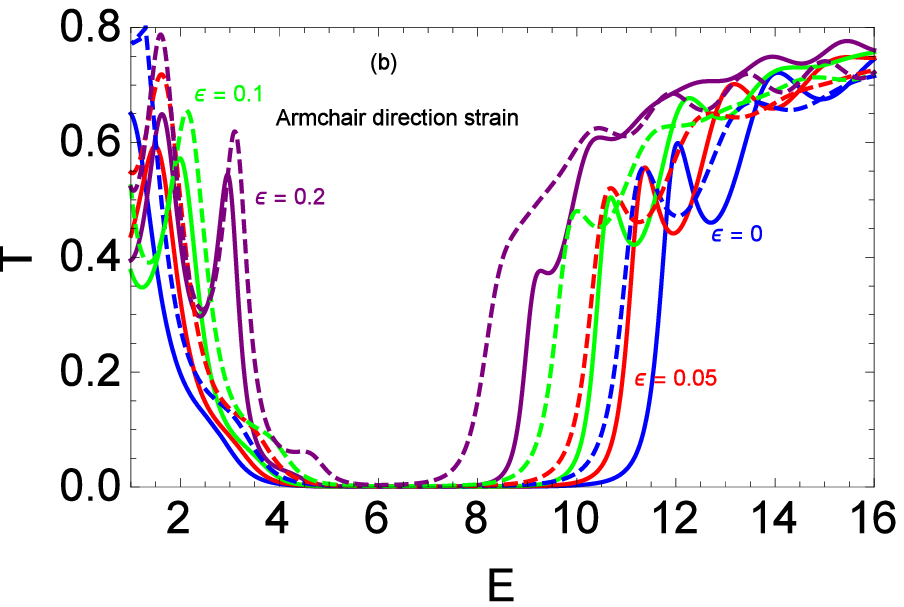}
\caption{\sf (color online) Transmission probability as a function of the energy
for different strains $\varepsilon=0.00, 0.05, 0.10,$ and
$0.20$. (a): Effects of zigzag  strain direction for up/down
spin $\eta=+/-1$. (b): The effects of armchair strain direction
for up/down spin $\eta=+/-1$ ($\eta=1$: solid line, $\eta=-1$:
dashed line). Other parameters are  $\gamma_\eta = 38^o, \omega =
0.8\ nm,$ $H_{ex} =0.61\ meV, V_b = 1.5\ meV, V_0 = 6\
meV,$ $l_B=0.5\ nm$. } \label{fig5}
\end{figure}

As we mentioned earlier, the strain can be applied 
along either the ZZ or AC directions. In Figure \ref{fig5} we show the
spin-dependent transmission probability as a function of the
energy for different strengths of the strain along the ZZ and AC
directions, respectively. We notice that when the deformation is
increased along the ZZ direction $(\theta=0)$ {or}
AC direction $(\theta=\frac{\pi}{2})$, it is found that the
zone in which the transmission is minimal can reach  zero at a
certain values of the energy (\emph{i.e.} transmission gap)
corresponding to  the forbidden band.
% and will therefore be
%widened.
Moreover, introducing a ZZ strain leaves the transmission profile
almost unchanged such that the associated peaks are located at
fixed energy. However, the overall transmission probability is
significantly reduced, see Figure \ref{fig5}(a). In addition, when
the strain follows the ZZ direction, the forbidden zone becomes
larger by increasing the strength of the strain. {In contrast}, when we
introduce a ZZ strain the transmission gap (\emph{i.e.} $T=0$)
becomes wider by decreasing the strength of the strain as shown in
Figure \ref{fig5}(a). {Moreover}, we notice that the peaks in the
spin up and down probabilities, for large $E$, are shifted in energy
by the exchange term value $\left\vert H_{ex} \right\vert $. For
example, the spin up ($\eta=1$) peak in the transmission
probability for the AC direction with $\varepsilon=0.1$ is situated at
$E\app10.7 \ meV$ and thus the counterpart peak in the spin down
($\eta=-1$) is located at $E-H_{ex}\app10.1\ meV$. On the other
hand, applying the AC strain to the sample completely alters the
transmission profile and increase the overall transmission, see
Figure \ref{fig5}(b). This is a manifestation of the different
band structure alignments along the carrier propagation direction
induced by the ZZ and AC strains.

\begin{figure}[!ht]
\centering
\includegraphics[width=8.3cm]{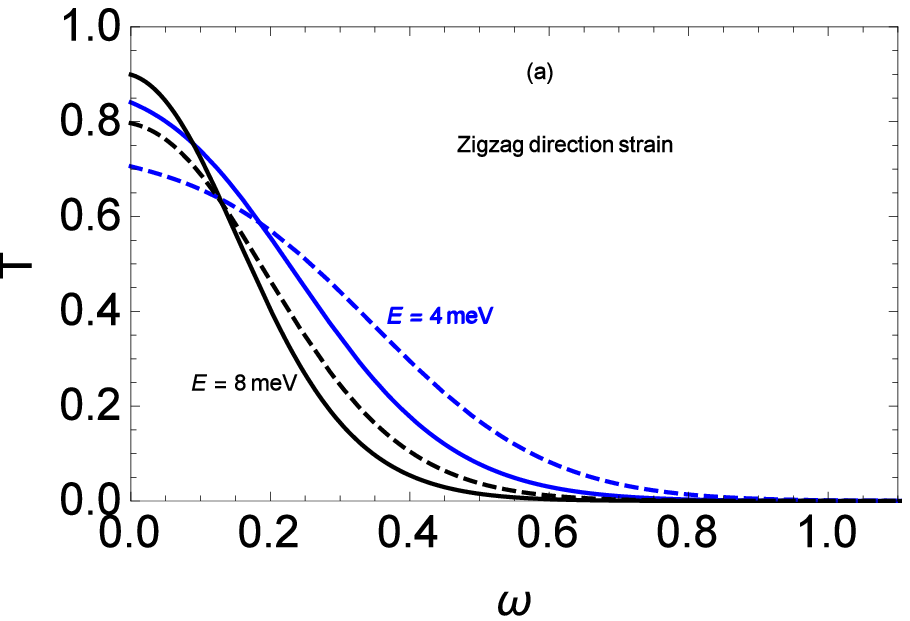}
\includegraphics[width=8.3cm]{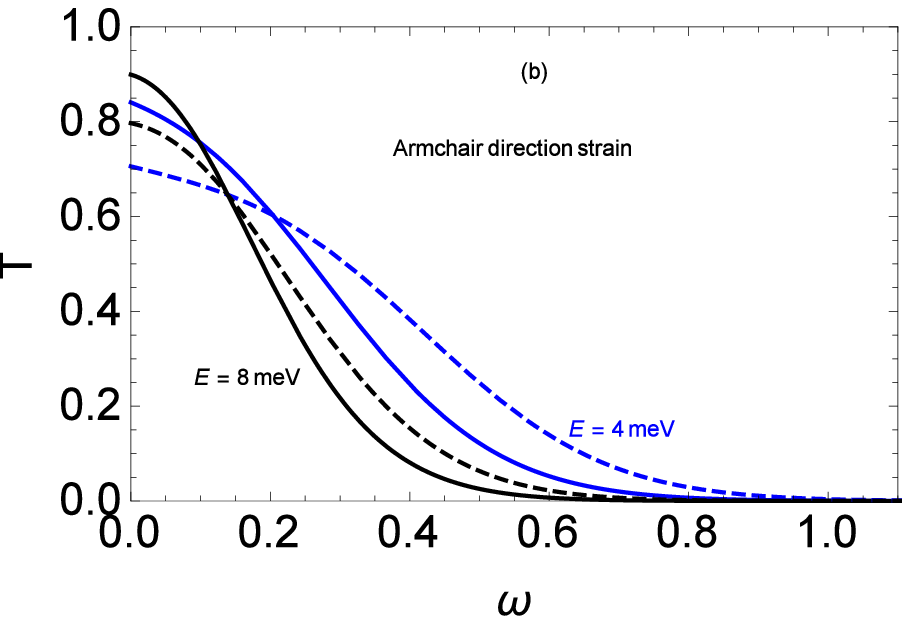}
\caption{\sf (color online) Transmission probability as a function of the width
of the central region $\omega$ for spin up $\eta=+1$ (solid line)
and spin down $\eta=-1$ (dashed line). (a)/(b): Effects of
zigzag/armchair strain direction. Other parameters are 
$\gamma_\eta = 38.1^{\circ}, H_{ex} = 0.61 \ meV, V_b = 1.50 \
meV, V_0=6 \ meV, \varepsilon=0.03, l_B=0.5\ nm$, $E=4\ meV$
(blue line) and $E=8\ meV$ (blue line).} \label{fig6}
 \end{figure}

Figure \ref{fig6} shows the transmission probability as a function
of the width $\omega$ of the strained region for up/down spin
and two values of the energy. It is clearly seen that, for a fixed
value of the energy, the transmission is maximal for a very thin
barrier. By increasing the width of the central region, the
transmission decreases until it reaches zero at a specific value of
$\omega$. It is important to note that the transmission, for the
two values of the energy, are nearly the same for both AC and ZZ
strains. On the other hand, the spin up transmission, for a thin
barrier, is larger than the spin down one. By increasing $\omega$,
%and from a specific value of the width,
the
situation will be reversed and the spin down transmission becomes
larger then the spin down one for both AC and ZZ strains.

 \begin{figure}[!ht]
  \centering
\includegraphics[width=8.3cm]{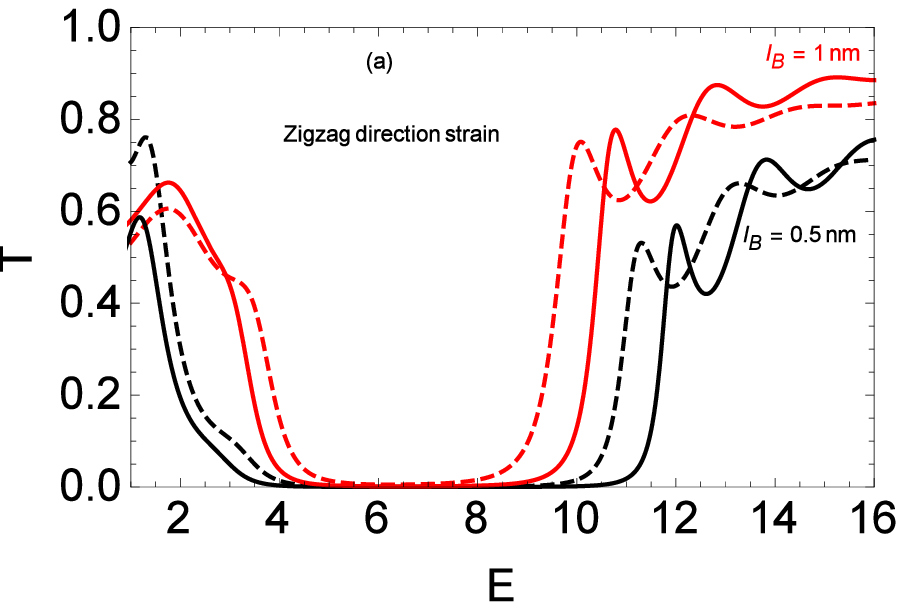}\
\includegraphics[width=8.3cm]{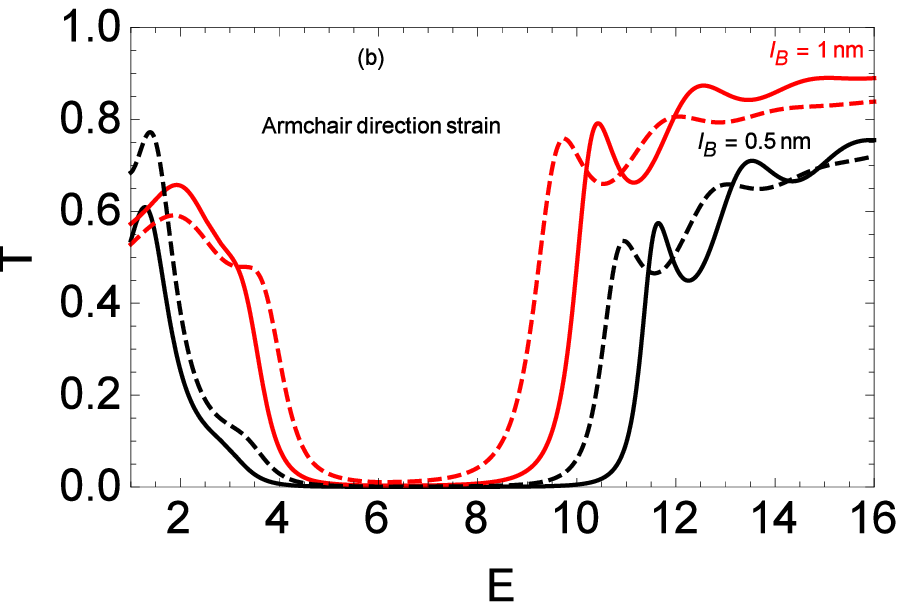}
\caption{\sf (color online) Transmission probability as a function of the energy
with different values of the magnetic length $l_B$ for spin
up/down $\eta=+/-1$ (solid/dashed lines). (a)/(b): Effects of
zigzag/armchair strain direction. Other parameters are 
 $\gamma_\eta = 38^{\circ}, \omega = 0.8\ nm, H_{ex} = 0.61\ meV, V_b = 1.50\ meV, V_0=6\
 meV$ and $\varepsilon=0.03$.} \label{fig7}
 \end{figure}

In Figure \ref{fig7}, we plot the transmission probability as a
function of the energy for different values of the magnetic
length. It is found that, like in Figure \ref{fig5}, the
transmission is minimal and reaches zero at certain values of the
energy, which  corresponds to a forbidden zone. Moreover, one
can clearly show that by increasing the magnetic field the
forbidden zone becomes wider. Note that the forbidden zone will
be widened if the deformation is along the AC direction. We
 observe that the forbidden zone corresponding to spin up
transmission is always {larger than that of  spin
down for both    ZZ and AC strains}. In addition, we see
clearly that some {peaks are shown up and their energy positions
vary with tuning the magnetic field}. We notice that the
transmission difference between spin up and spin down  is
{small} before the forbidden zone. However, the difference becomes
more important for large values of the energy.

 \begin{figure}[!ht]
  \centering
\includegraphics[width=8.3cm]{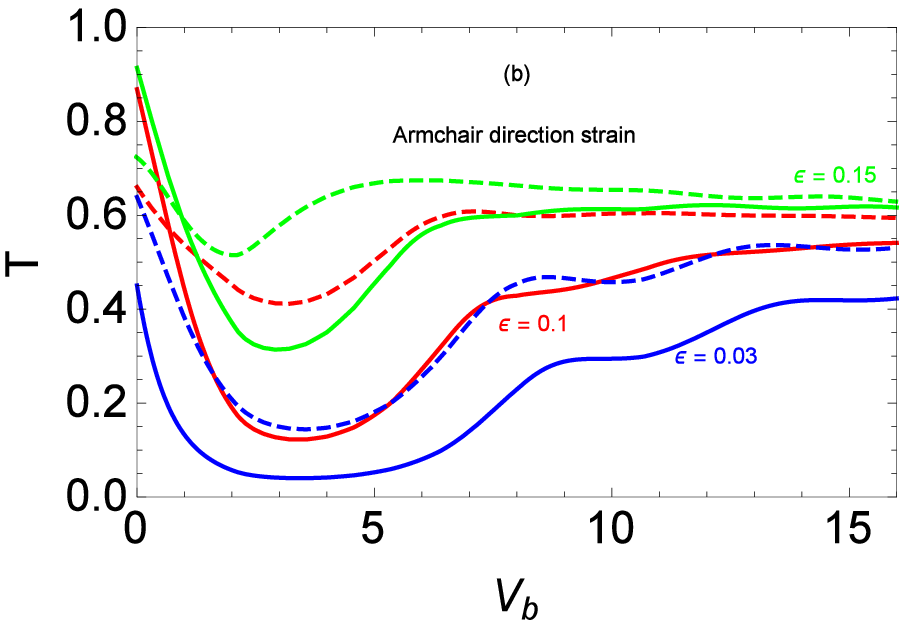}\
\includegraphics[width=8.3cm]{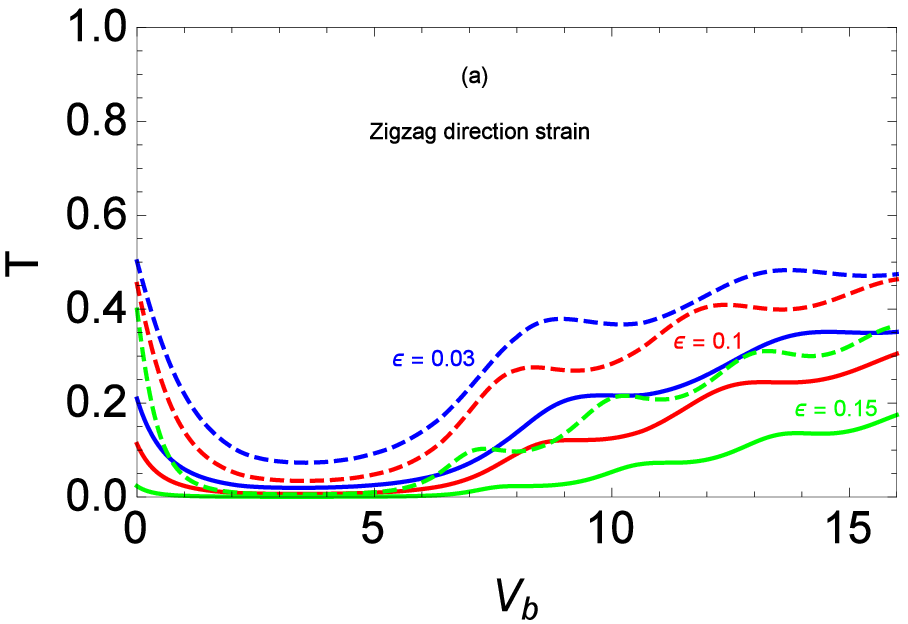}
\caption{\sf (color online) Transmission probability as a function of the bias
$V_b$ with different strains $\varepsilon=0.03$ (blue line),
$\varepsilon=0.1$ (red line) and $\varepsilon=0.15$ (green line)
for spin up/down $\eta=+/-1$ (solid/dashed lines). (a)/(b): 
Effects of zigzag/armchair strain directions with
%The other parameters
%are : 
$\gamma_\eta = 38^{\circ}, \omega = 0.8\ nm, H_{ex} =
0.61\ meV, E = 4\ meV, V_0=6\ meV$, $l_B= 1 \ nm$.}
\label{fig8}
 \end{figure}

In Figure \ref{fig8}, we investigate the transmission as a
function of the bias $V_b$ for different strain magnitudes. It
is clearly shown that, for $V_b=0$ we have a transmission that
{decreases} until it {reaches} a minimal value by increasing the bias. For
large values of the bias, the transmission starts increasing and
{exhibits}  some {oscillations} that are, then, relatively damped. It
is important to note that the spin up and down transmission have
almost the same form, but the spin down transmission is always
 larger. For the AC strain, we  see that the transmission
{increases} up to a specific value and then reaches  almost a constant value.
Moreover, when the bias $V_b$ is near  zero (for the AC strain) the
transmission is maximal, which is {not the case} for the ZZ strain.

%%%%%%%%%%%%%%%%%%%%%%%%%%%%%%%%%%%%%%%%%%
\section{Conclusion}
%%%%%%%%%%%%%%%%%%%%%%%%%%%%%%%%%%%%%%%%

We have investigated  the transport properties of a
ferromagnetic/strained/normal graphene junctions where 
a magnetic field is applied in the central region. After writing down the corresponding
Hamiltonian, the eigenvalue
equation has been solved in each region 
composing our system to end up with the solutions of the energy spectrum. %eigenspinors and associated eigenvalues.
%,in the three
%regions were obtained as function of the system parameters.
These have been used together with
%Using 
the transfer matrix approach and density current to determine the transmission and reflection
probabilities  in terms of the physical parameters
characterizing our system.
%matching the eigenspinors at interfaces we have calculated the
%corresponding transmission {probabilities}.

Our numerical results showed that in the first region ({\sf I}) the spin up
and down bands are shifted by an amount equal to the exchange energy
$H_{ex}$ and left the whole spectrum linear as in the case of
pristine graphene. However, in the third region ({\sf III}), it is {found} that
the energy bands are still linear like in the case of pristine
graphene, but shifted down by $V_b$. Moreover, by changing the
values of the applied magnetic field, the position of the Dirac
point changes due to the uniaxial
strain. In the strained region ({\sf II}) 
%the
%story is completely different due to the deformation of the
%graphene sheet and the applied magnetic field. In fact, 
we have showed that the
application of deformation leads to a renormalization of the Fermi
velocity, \emph{i.e.} $v^{*}_F=\xi v_F$ and  the
magnetic field leaves the energy spectrum quantized,
which results in the so-called Landau Levels (LLs). As a results, we have {found}
that due to the applied strain, the LLs are contracted with
respect to the pristine graphene. Our numerical results showed
also that by increasing the strain magnitudes, under the same
magnetic field, both the energy and distance between the
LLs decrease, which is due to the fact that the applied strain
affects the cyclotron orbital motion.

We have studied the transmission probabilities, for two
directions of strain including zigzag (ZZ) and armchair (AC), as a function of
the incident energy,  width of the central region and {electrostatic grate}
$V_b$. It is found that the transmission {exhibits}  a forbidden zone
($T=0$), at a certain values of the energy, when the
strain deformation is {along   either  the ZZ or AC directions}. In
addition, the width of this zone {increases} by increasing the
strength of the strain {along the  ZZ direction} but  {decreases}
for the AC strain. The situation was quite different 
{by tuning the  applied magnetic field} because the width of the forbidden
zone {increases} with decreasing  magnetic field for {either the ZZ or AC}
strain directions. Therefore, some peaks occur in the transmission
and their energy position {are changed} by {altering} the strain magnitude.
Moreover, we have observed that, for {low} energy, the difference
between the spin up and down transmission is small before the
forbidden zone and becomes more important after that. On
other hand, we have {found} that by increasing the width of the
strained region the transmission {decreases} {and  reaches zero}.
Moreover, the transmission {increases} to reach a minimal value for
{small} values {of the electrostatic gate } $V_b$ and  as long as  $V_b$ is increased the
{transmission} {increased} up to a specific value and
reached almost constant. The spin up and down transmission have
almost the same form, but the spin down transmission is almost {always
 larger}.

%%%%%%%%%%%%%%%%%%%%%%%%%%%%%%%
\section*{Acknowledgments}
%%%%%%%%%%%%%%%%%%%%%%%%%%%%%%

The generous support provided by the Saudi Center for Theoretical
Physics (SCTP) is highly appreciated by all authors. AJ
and HB
acknowledge the support of King Fahd University of Petroleum and
minerals under research group project RG171007.

\end{document}